\documentclass[11pt]{amsart} 
 
\usepackage{latexsym} 
\usepackage{amsfonts} 
\usepackage{amsmath} 
\usepackage{amsthm} 
\usepackage{mathrsfs} 
\usepackage{dsfont} 
\usepackage{bbold} 
\usepackage[english]{babel} 
\usepackage{caption} 
\usepackage{epsfig} 
\usepackage{float} 
\usepackage{subfigure} 
\usepackage{psfrag} 
\usepackage{graphicx} 
\usepackage{epsfig} 
\usepackage{amsbsy,esint} 
\usepackage{hyperref}

\newtheorem{theorem}{Theorem} 
\newtheorem*{prop*}{Theorem} 
\newtheorem{theo}[theorem]{Theorem} 
 
\newtheorem{defi}[theorem]{Definition} 
\newtheorem{lemma}[theorem]{Lemma} 
\newtheorem{mlemma}[theorem]{Meta-Lemma} 
\newtheorem{prop}[theorem]{Proposition} 
\newtheorem{rmk}[theorem]{Remark}

\newcommand{\zerarcounters}{\setcounter{equation}{0}\setcounter{theorem}{0}} 
 
\newcommand{\beq}{\begin{equation}}
\newcommand{\eeq}{\end{equation}}

\newcommand{\ZZZ}{\mathds{Z}} 
\newcommand{\CCC}{\mathds{C}}

\newcommand{\RRR}{\mathds{R}} 
\newcommand{\TTT}{\mathds{T}} 
\newcommand{\uno}{\mathds{1}} 
 
\newcommand{\calA}{{\mathcal A}} 
 
\newcommand{\CCCC}{{\mathcal C}}

\newcommand{\Fullbox}{{\rule{2.0mm}{2.0mm}}} 
 
\newcommand{\EP}{\hfill\Fullbox\vspace{0.2cm}} 
\newcommand{\prova}{\noindent{\it Proof. }} 
\newcommand{\io}{\infty} 
\newcommand{\e}{\varepsilon} 
\newcommand{\al}{\alpha} 
\newcommand{\de}{\delta} 
\newcommand{\be}{\beta} 
 
\newcommand{\m}{\mu} 
\newcommand{\x}{\xi}

\newcommand{\g}{\gamma} 
\newcommand{\om}{\omega} 
\newcommand{\h}{\eta} 
 
\newcommand{\la}{\lambda} 
\newcommand{\f}{\varphi}

\newcommand{\del}{\partial} 
\newcommand{\laa}{\langle} 
\newcommand{\raa}{\rangle}

\newcommand{\avg}[1]{\langle #1 \rangle}

\def\tilde#1{\widetilde{#1}}
 
\def\ins#1#2#3{\vbox to0pt{\kern-#2 \hbox{\kern#1 #3}\vss}\nointerlineskip} 
 
\begin{document}
 
\title[Expansions for SDDEs]{Expansions in the delay  of quasi-periodic solutions 
for state dependent delay equations}

\author[A. Casal]{Alfonso Casal} 
\address{Dept. Matem\'atica Aplicada \\ 
E.T.S.A. Universidad Polit\'ecnica de Madrid\\
Av. Juan de Herrera 4, Madrid 28040}
\email{alfonso.casal@upm.es}
 
\author[L. Corsi] {Livia Corsi}
\address{Dip. di Matematica e Fisica \\ Universit\`a di Roma Tre \\ 
L.go S. Leonardo Murialdo 1 \\ 00146 Roma, Italy} 
\email{lcorsi@mat.uniroma3.it} 

\author[R. de la Llave]{Rafael de la Llave}
\address{Dept. of Mathematics \\ Georgia Inst. of Technology\\ 
686 Cherry St. \\ Atlanta, GA 30332-0160}
\email{rll6@math.gatech.edu}
\thanks{Supported in part by NSF grant DMS 1800241}

\begin{abstract} 
We consider several models of State Dependent Delay Differential Equations (SDDEs), in 
which the delay is affected by a small parameter.   This is a 
very singular perturbation since the nature of the equation changes. 

Under some conditions, we construct formal 
power series, which solve the SDDEs order by order.
These series are quasi-periodic functions of time. 
This is very similar to  the Lindstedt procedure in celestial mechanics. 

Truncations of these power series can be taken as input
for a-posteriori theorems, that show that near the approximate solutions there are 
true solutions.  
In this way, we hope that one can construct 
a catalogue of solutions for SDDEs, bypassing the need of 
a systematic theory of existence and uniqueness for all initial conditions. 
\end{abstract} 

\maketitle 
  
 \tableofcontents

\zerarcounters 
\section{Introduction} 
\label{intro} 

 In special relativity, when we can ignore the effect of emitted radiations,
\cite{WheelerF49, Rohrlich, LandauL, Spohn}, 
 the motion of $N$ charged particles 
 can be described as solutions of the equation 
 \begin{equation}\label{dynamics}
  M_i(\dot{x}_i) \ddot{x}_i = G \sum_{j\ne i} \frac{q_i q_j  (x_i(t) - x_j(\tau_{ij}))}{|x_i(t) - x_j(\tau_{ij})|^3}\,,
 \end{equation}
 where $M$ is the relativistic mass, which in the one dimensional case is defined as
$$
 M_i(v) = m_i \left(1-\frac{|v|^2}{c^2}\right)^{-3/2} + m_i \left(1-\frac{|v|^2}{c^2}\right)^{-1/2}\,,
$$
$m_i$, $q_i$ are the rest  mass and the charge of the $i$-th particle respectively, $c$ is the speed of light and $G$ is a physical constant.
(In the higher dimensional case, the relativistic mass is a matrix). 
More importantly,
 $\tau_{ij}$ is the time that a signal emitted from particle 
$j$ takes to reach particle $i$, and it is given by the implicit equation
 \begin{equation}\label{tau}
\tau_{ij} = t- \frac{1}{c} |x_i(t) - x_j(\tau_{ij})|\,.
\end{equation}  
It is not difficult to show that if all the particles move with 
speed less that the speed of light, the solution of \eqref{tau} is a unique function of 
$t$.

In Physics, it is common to consider 
$\e \equiv \frac{1}{c}$ as a small parameter
and to try to predict the motion as a formal power series; see for instance \cite{MTW}.
It is important to observe that, if we
consider $x_i$ as given, we can find an asymptotic expansion 
of the solutions of \eqref{tau}. Indeed we can 
write 
\begin{equation}\label{tauexpansion} 
\tau_{ij} =  t - \frac{1}{c}| x_i(t) - x_j(t)| + O(1/c^2)
\end{equation}

The functional equation \eqref{dynamics} is not a differential
equation because $\tau_{ij}$ is in general not equal to $t$, hence the
positions $x_i$ in the r.h.s. are evaluated at different times.
If we take the approximation \eqref{tauexpansion} for 
the delay, we obtain a State Dependent Delay Equation (SDDE) 
because the delay is a explicit function of 
the state.  The problem in \eqref{dynamics}, 
without making the simplification \eqref{tauexpansion} is a  
of a more complicated nature since 
$\tau_{ij}$ depends implicitely on the whole 
trajectory of $x_i$, $x_j$. After we develop enough 
theory for SDDE, we will see in Section~\ref{sec:electrodynamics} 
that the same ideas apply to the full model.

Other scientific problems, such as the dynamics in some population
models with density dependent fertility age, are also naturally
modeled with SDDEs; \cite{HKWW06,KNbook,IM}. 

\medskip

In this paper we will
study several models of delay equations (mainly State dependent delay 
equations. 

Note that when the delay is a given constant $T$, there is a rather 
developed mathematical theory \cite{Hale77,HaleVL93,DieckmannGVL95}. Precisely, if 
one prescribes as initial data a function defined on $[0,T]$, under 
the standard regularity assumptions for classical ODEs, one can obtain a rather satisfactory
theory of existence, uniqueness, dependane on parameters and initial data, which constitutes the first step 
to developing a qualitative theory. 
However, when the delay depends on the state of the system
(a fortiori on the whole trajectory)  the situation is 
much more delicate and the theory of existence and uniqueness 
is much more restricted \cite{HKWW06}. There are indeed many examples of surprising
behaviors which indicate that a systematic existence, uniqueness and 
regularity theory for SDDE will be significantly more complicated than the one for constant delay equations.

%

The goal of this paper is mathematically modest.  We do not try 
to develop a general 
 theory of existence and uniqueness. We only try to study 
special solutions for some type of SDDEs. Furthermore, 
we only try to study these solutions as formal power series. 

Once we specify the class of solution we are looking for 
(mainly quasi-periodic\footnote{In  Section~\ref{sec:limitcycle} 
we will consider also solutions converging exponentially to 
quasi-periodic}) 
we express
the SDDEs as functional equations on the space of quasi-periodic functions, and we call such
functional equation the \emph{invariance equation}. 
In particular, we will consider SDDEs involving a small parameter $\e$, and
obtain approximate solution of the invariance equation as a formal power series in $\e$.

One motivation for 
our study is that quasi-periodic solutions of \eqref{dynamics} play 
an important r\^ole in chemistry since
they are the basis for the ``old quantum theory''.
For $\e=0$, the quasi-periodic solutions of \eqref{dynamics} satisfying the Bohr-Sommerfeld conditions
have quantum analogues. 

This is the background we mainly have in mind, and thus it motivates us to
look for 
expansions in $\e=1/c$ of quasi-periodic solutions for \eqref{dynamics}.

Thus, we will show that, under some mild non-degeneracy condition, it is possible 
to write systematically a formal power series expansions for the quasi-periodic solutions. 
Furthermore we will show that, if we trucate such expansions to a finite order, we obtain 
functions that, when substituted in the invariance equation,
 satisfy it up to a very small error. 

Note that this is a very different procedure from the one 
followed sometimes in the Physics literature (predictive mechanics, 
\cite{Bel, MS}
Postnewtonian formalism \cite{MTW} etc. ) in which one tries to find 
an ODE (often a derived through a Lagrangian) which describe all 
the solutions. Our aim is to find expansions only for 
solutions of a certain type. It is quite possible, with the formalism 
developed here, that the perturbation expansions for solutions of 
different types are very different.

  Note that obtaining a Lagrangian description of 
the motion of all particles, is forbiden by the ``no-interaction'' theorems
\cite{CJS}, which state that the only Lagrangian invariant 
under the Lorentz transformations are the free particles. Indeed, in general
not even formal power series can be found \cite{H}. 
The above results are not incompatible
with our results, since we obtain the expansion only for solutions of 
very specific type. As observed in Remark~\ref{liou}, our 
expansions depend very much in subtle properties of 
the unperturbed system, so it is quite possible that the approximate 
solutions we produce cannot be combined into a globally defined 
Lagrangian sytem, which is the only thing forbidden by \cite{CJS}. 

The systematic construction of  approximate solutions obtained in this paper
matches very well with the recent developments in a-posteriori 
theorems, which show that near approximate solutions of a certain 
kind there will be true solutions.   There are already 
such a-posteriori results in quasi-periodic perturbations of 
some simple systems \cite{HE1, HE2} and in 
\cite{YangGL19, GimenoYL19}. Putting together  
these results, we obtain that some of the expansions we 
construct are asymptotic expansions of families of true solutions. 
One can hope that in the near future, the (rapidly growing) 
 applicability of 
a-posteriori theorems will be extended and more general theorems 
of this form will be proved, to cover at least the models considered 
in this paper. We call attention to \cite{Chicone03} which implemented 
a very similar program of finding expansions in the delay and validating them.

Hence, the conjectural picture that emerges is that there 
are many solutions of the classical system that survive 
the inclusion of the delay. The set of solution that persists 
has a very complicated structure (number theory properties
play a role) and the solutions depend in a very non-uniform way. 
There is no way to make the solutions that persist fit in a
common Lagrangian description, but nevertheless, the set of
solutions that persist is large enough that they can be useful 
in practical problems. 

\medskip

\noindent
{{\it Acknowledgements}}. We  thank C. Chicone, S. Dostoglou, J. Gimeno and J. Yang for discussions and encouragement. 
L.C. wishes to thank the School of Mathematics of Georgia Institute
of Technology  for the nice hospitality.

 \subsection{Formulation of the problem}\label{problema}
We consider equations of the form
\begin{equation}\label{gene}
\dot y(t) = f_\e (y(t), y(t-\e r_1),\ldots, y(t - \e r_{\ell}))\,,
\end{equation} 
or 
\begin{equation}\label{vabbeh}
\dot y(t) = f_\e (y(t), \e y(t- r_1),\ldots, \e y(t - r_{\ell}))\,,
\end{equation} 
where $r_j=r_j(y(t))$, $j=1,\ldots,\ell$ are given functions, and the unknown is $y(t)$.
In \eqref{dynamics} the small parameter is $\e=1/c$. 

\begin{rmk}\label{ezero}
 For $\e=0$ the resulting equation is an ODE in both cases \eqref{gene} and \eqref{vabbeh}.
\end{rmk}

\begin{rmk}
We can think of \eqref{dynamics} as an equation of the form
\begin{equation}\label{genebis}
\dot y(t) = f_\e (y(t), y(\tau))\,,
\end{equation} 
with $y\in \RRR^{6N}$ (positions and velocities), $\tau=\tau(y(t))=\{\tau_{i,j}\}_{i,j=1}^N$ is implicitely defined by \eqref{tau},
and $\ell = N(N-1)/2$ is the number of pairs. In particular, for $\e=0$ the equation \eqref{dynamics} has an Hamiltonian structure.
\end{rmk}

For the sake of typographical simplicity, in this paper we will present mostly cases in which $\ell=1$, and at the end we will make explicit the (typographical)
changes needed to deal with the case $\ell\ge2$ or the more complicated model of \eqref{dynamics}.

\medskip

The search of quasi-periodic solutions of \eqref{gene} with some frequency $\om$
is equivalent to looking for a so-called \emph{invariant torus}, i.e. a torus embedding
\begin{equation}\label{cappone}
K:\TTT^d\to \RRR^n
\end{equation}
satisfying
\begin{equation}\label{embe}
(\om\cdot\del_\theta K)(\theta) = f_\e (K(\theta),K(\theta - \e \om r(K(\theta)))),
\end{equation}
in such a way that the dynamics on the model torus $\TTT^d$ is given by
\begin{equation}\label{costante}
\dot \theta = \om\,.
\end{equation}
Of course if dealing with \eqref{vabbeh}, we look for $K$ satisfying
\begin{equation}\label{embe2}
 (\om\cdot\del_\theta K)(\theta)= f_\e (K(\theta), \e K(\theta -  \om r(K(\theta))))\,.
\end{equation}

Observe that the case $d=1$ corresponds to periodic solutions.
Note also that 
\eqref{embe} reduces to 
\begin{equation}\label{zero}
(\om\cdot\del_\theta K)(\theta) = f_0 (K(\theta), K(\theta)),
\end{equation}
for $\e=0$, while \eqref{embe2} reduces to
\begin{equation}\label{zer2}
(\om\cdot\del_\theta K)(\theta) = f_0 (K(\theta), 0).
\end{equation}

We emphasize that in \eqref{embe} and \eqref{embe2} both $K$ and $\om$ are unknown. In \cite{HE1,HE2} 
only the simpler case of quasi-periodically forced systems was considered, so that $\om$ was externally fixed.

\begin{rmk}\label{uni}
Note that the solutions of \eqref{embe} or \eqref{embe2} are never unique. Indeed if $K_\e$ is a solution, for any $\theta\in\RRR^d$
we have that $\tilde{K}_\e(\theta) := K_\e(\theta+\om)$ is also a solution, i.e. we may say that the solution admits
phase traslations, i.e. traslating the origin in $\TTT^d$.  A way to obtain uniqueness is by requesting,
besides the invariance, a normalization: the most natural one seems to be
\begin{equation}\label{media}
\frac{1}{(2\pi)^d}\int_{\TTT^d} d\theta DK_0(\theta) K_\e(\theta) =0\,.
\end{equation}

Having the uniqueness of the solution is a useful property since it allows to compare results obtained by
different methods.
\end{rmk}

The theory for the solutions of \eqref{embe} is far from being a general theory for the solutions of \eqref{gene} for all initial data.
The goal of this paper is to show that if we have $K_0$ and $\om_0$ solving \eqref{embe} for $\e=0$
and we assume some mild non-degeneracy conditions, then we can systematically compute formal power series
\begin{equation}\label{series}
K_\e = \sum_{j\ge0} \e^j K_j \qquad\qquad
\om_\e = \sum_{j\ge0} \e^j \om_j
\end{equation}
solving \eqref{embe} in the sense of formal power series. In other words, if we denote
\begin{equation}\label{tronco}
K_\e^{[\le N]} = \sum_{j=0}^N \e^j K_j \qquad\qquad
\om_\e^{[\le N]} = \sum_{j=0}^N \e^j \om_j
\end{equation}
we have that the function
\begin{equation}\label{appro}
y_\e^{[\le N]}(t) := K_\e^{[\le N]}(\om_\e^{[\le N]}t)
\end{equation}
satisfies
\begin{equation}\label{taglio}
\left|
\frac{d}{dt}y^{[\le N]}_\e - f_\e (y^{[\le N]}_\e(t), y^{[\le N]}_\e(t-\e r(y^{[\le N]}_\e(t)))
\right| \le C_N \e^{N+1}\,.
\end{equation}

For \eqref{embe2} the analogue of \eqref{taglio} is

\begin{equation}
\left|
\frac{d}{dt}y^{[\le N]}_\e - f_\e (y^{[\le N]}_\e(t), \e y^{[\le N]}_\e(t- r(y^{[\le N]}_\e(t)))
\right| \le C_N \e^{N+1}\,.
\end{equation}

Expansion of the form \eqref{series} are called ``Lindstedt series'' and the search of solutions for an ODE
in the form of a Lindstedt series has been widely used in astronomy since the 19-th century \cite{Poi} and even before. 
Such expansions have been used also in delay equations; see for instance \cite{CF,CF2} or \cite{Rand} for further developement. The paper \cite{Chicone03} 
includes also validation.

An important r\^ole will be played by the linearized equation around $\e=0$.
Postponing many details which we will make explicit later, a key result of this paper is the following meta-result.

\begin{mlemma}\label{formal} Denote by  $D_1,D_2$ the derivative w.r.t. the first and second argument of $f_0$ respectively.
If given $R$ it is possible to find
$\de$ ``small enough'' and $u$
such that
\begin{equation}\label{tengo}
\om_0\cdot\del_\theta u - \Big(D_1f_0(K_0(\theta),K_0(\theta)) + D_2f_0(K_0(\theta),K_0(\theta))\Big)  u = R + \de \del_\theta K_0(\theta )\,,
\end{equation}
then we can determine the coefficients of the series \eqref{series} solving \eqref{embe} to all orders. 

Similarly, if
it is possible to find
$\de$ ``small enough'' and $u$
such that
\begin{equation}\label{tengo2}
\om_0\cdot\del_\theta u - \Big(D_1f_0(K_0(\theta),0) + D_2f_0(K_0(\theta),0)\Big)  u = R + \de \del_\theta K_0(\theta )\,,
\end{equation}
then we can determine the coefficients of the series \eqref{series} solving \eqref{embe2} to all orders. 
\end{mlemma}

It is important to note that the the equations
\eqref{tengo} and \eqref{tengo2} only involve the unperturbed undelayed 
equation  
and that the same equations appear to all orders. Hence, some 
geometric properties of the solutions of the ODE, guarantee that we 
can get expansions to all orders in $\e$. 

\prova
The proof is quite straightforward and we shall show the details only for the case \eqref{embe}. We start by simply observing that 
\[
\begin{aligned}
f_\e(K_\e(\theta), K_\e(\theta-\e\om r (K_\e(\theta)))) = &f_0(K_0,K_0) + \e D_1 f_0( K_0(\theta),K_0(\theta)) \cdot K_1(\theta) \\
&\qquad+\e D_2 f_0( K_0(\theta),K_0(\theta)) \cdot K_1(\theta)+\ldots
\end{aligned}
\]

Thus, by a formal expansion in $\e$, we see that the terms $O(\e^n)$ have the form
\begin{equation}\label{highor}
\begin{aligned}
(D_1f_0 (K_0,&K_0) +D_2f_0(K_0,K_0))  K_n \\
&+ R_n(K_0,\ldots,K_{n-1},DK_0,\ldots,DK_{n-1},\ldots, D^{n-1}K_0,\ldots, D^{n-1}K_{n-1})
\end{aligned}
\end{equation}
where $R_n$ is a polynomial in its variables, i.e. matching the coefficients at order $\e^n$
both in $K$ and in $\om$ we obtain an equation of the form \eqref{tengo}.
\EP

Of course the statement of Meta-Lemma \ref{formal} above is only formal since it does not specify the precise meaning of ``solve''.
Such precise meaning entails the specification of the spaces in which the solution and the reminder lie; moreover
we will need conditions on the frequency $\om_0$.

\medskip

In the following we will present various cases in which the equation \eqref{tengo} (or \eqref{tengo2}) is solvable. For each of the cases
we formulate precisely the meaning of ``solvability'' and the result on the existence of the Lindstedt series.
The cases we will consider are well known to dynamicysts since they are also cases where one can prove 
persistence of the structure under the change of the differential equation, and they are the following.

\begin{itemize}

\item[{\bf Case 1}] The manifold $K_0(\TTT^d)$ is a Normally Hyperbolic Invariant Manifold (NHIM) and $\om_0$ satisfies a Diophantine condition.

\item[{\bf Case 2}] The linearized evolution is reducible to constant coefficients and the eigenvalues of this constant coefficients matrix satisfy 
some Diophantine condition w.r.t. $\om_0$.

\item[{\bf Case 3}] The unperturbed system is Hamiltonian, $K_0(\TTT^d)$ is a Lagrangian torus (i.e. the phase space has dimension $2d$), it satisfies a twist
condition and $\om_0$ satisfies some Diophantine condition.

\item[{\bf Case 4}] The unperturbed system is a two dimensional 
ODE which  has a limit cycle.

\item[{\bf Case 5}] The ``electrodynamics case'' of \eqref{dynamics}
(The delay depends not only on the state, but also on the whole trajectory).

\end{itemize}

Here and henceforth we impose the standard Diophantine condition
\begin{equation}\label{dio}
|\om_0\cdot k| \ge \frac{\g}{|k|^\tau},\qquad \mbox{ for all }k\in\ZZZ^d\setminus\{0\}\,,
\end{equation}
where, with abuse of notation, we denoted by $|\cdot|$ both the absolute value of a number and the $\ell_1$-norm of a $d$-dimensional vector.

Some result can be obtained also in the case of  {\it subexponential Diophanitne} $\om_0$, i.e.
\begin{equation}\label{subexp}
\lim_{|k|\to\io} \frac{1}{|k|}\log\frac{1}{|\om_0\cdot k|} =0
\end{equation}
or equivalently
$$
\forall\ \e>0 \ \exists \ c=c(\e) {\mbox{ such that }} |\om_0\cdot k| \ge c(\e)e^{-|k|\e}\qquad\forall\ k\in\ZZZ^d\setminus\{0\}\,.
$$

It is remarkable that the Diophantine condition \eqref{dio} is precisely one of the two main hypotheses
of the celebrated KAM theorem.

Throughout the paper we shall use the following standard notations.

\begin{itemize}

\item Given $\x>0$ we denote by $\TTT^d_\x$ the set
$$
\TTT^d_\x = \{\theta\in (\CCC/\ZZZ)^d\;:\; \mbox{Re}(\theta_j)\in\TTT,\ |\mbox{Im}(\theta_j)|<\x,\ j=1,\ldots,d\}\,.
$$

\item We denote by $\calA_\x$ the space of functions $u:\TTT^d_x\to\RRR^n$ such that
\begin{equation}\label{ax}
\|u\|_{\x} := \sum_{k\in\ZZZ^d} e^{\x2\pi|k|}\|\hat{u}_k\| <\io
\end{equation}
where we denoted by $\hat{u}_k$ the $k$-th Fourier coefficient of $u$ and by $\|\cdot\|$ the standard Euclidean norm of an $n$-dimensional vector.

\item For a function $f$ of class $C^r$ we denote its $C^r$-norm as $\|f\|_{C^r}$.

\end{itemize}


\zerarcounters 
\section{The case of  of quasi-periodic solutions 
which are also normally hyperbolic invariant manifolds} 
\label{nhim} 

\subsection{Basic definitions}

We recall that $M\subset \RRR^n$ a $C^1$ manifold is a NHIM for a $C^r$ vector field $f$ with $r\ge2$ if

\begin{enumerate}

\item $f(x)\in TM$ for all $x\in M$

\item For every $x\in M$ there is a splitting 
\begin{equation}\label{split}
\RRR^n=T_xM\oplus E^s_x\oplus E^u_x,
\end{equation}
such that there are positive constants $C$, $\rho_+$, $\rho_-$ so that,
denotig by $F_t$ the time-$t$ flow, one has
\[
\begin{aligned}
&\|{DF_t}_{|_{E^u}}\|_{C^r} \le C e^{\rho_- t},\qquad t\le0\\
&\|{DF_t}_{|_{E^s}}\|_{C^r} \le C e^{-\rho_+ t},\qquad t\ge0
\end{aligned}
\]

\end{enumerate}

Note that if $M$ is not compact, one needs to assume that the $C^r$ properties of the manifold are uniform, 
and in this case the theory of \cite{Fenichel, HPS, Pesin04} carries through; see also \cite{BLZ}. 

Here $M$ is $K_0(\TTT^d)$ which is a compact manifold, so that we do not have to deal with the subtleties
appearing in the case of non-compact manifolds.

We assume that
 the dynamics restricted to $M$ is conjugated to a rotation. 
In such a case, the theory of \cite{Fenichel} shows that if $f_0\in C^r$, then $M$ is a $C^r$ submanifold and the splitting is $C^{r-1}$.
The analytic case is more delicate, but it was proved in \cite{CH} that if $\om_0$ is Diophantine and $f_0$ is analytic,
then $M$ and the splitting are also analytic.

\subsection{Solvability of the linearized equation}

Denote by $\Pi^s_x$, $\Pi_x^u$, $\Pi^c_x$ the projectors onto $E^s_x$, $E^u_x$, $E^c_x=T_xM$ corresponding to the
splitting \eqref{split}.

Denoting
\begin{equation}\label{linh}
\begin{aligned}
&v_\theta^\al := \Pi_\theta^\al v,\\
&F^\al(K_0(\theta)) := \left.(D_1f_0(K_0(\theta),K_0(\theta)+D_2f_0(K_0(\theta),K_0(\theta))\right|_{E^\al_\theta},\qquad \al=s,u,c,
\end{aligned}
\end{equation}
the linearized equation \eqref{tengo} takes the form 
\begin{equation}\label{zato}
\begin{aligned}
&\om_0\cdot \del_\theta v^\al - F^\al( K_0(\theta)) v^\al = R^\al, \qquad \al=s,u\\
&\om_0\cdot \del_\theta v^c - F^c(K_0(\theta)) v^c= R^c  -\om_n \,.
\end{aligned}
\end{equation}
We need to show how to solve \eqref{zato}.

First of all we note that
for $\al=s,u$ it sufficies to use the Duhamel formula. Indeed if $A_\theta^\al(t)$ satisfies 
\begin{equation}\label{aone}
\left\{
\begin{aligned}
&\frac{d}{dt}A^\al_\theta (t) = F^\al(K_0(\theta +\om_0 t)) A^\al_\theta(t) \\
&A_\theta^\al (0) = \uno
\end{aligned}
\right.
\end{equation}
then we can set
\begin{equation}\label{int}
v^s(\theta) = \int_0^\io dt A^s_{\theta-\om_0 t}(t) R^s({\theta-\om_0 t})\,,\qquad \qquad
v^u(\theta) = \int_{-\io}^0 dt A^u_{\theta-\om_0 t}(t) R^u({\theta-\om_0 t})\,.
\end{equation}
Since we have
$$
|A^s_\theta(t)| \le C e^{- \rho_+ t},\quad \forall\ t>0\qquad\qquad
|A^u_\theta(t)| \le C e^{\rho_- t},\quad \forall\ t<0
$$
then the integral appearing in \eqref{int} is convergent, so $v^\al(\theta)$ is well-defined.

\begin{lemma}\label{lem.sol}
If $f_0\in C^r$ and $R^\al\in C^{r-1}$ for $\al=s,u$, then $v^\al\in C^{r-1}$ and one has
\begin{equation}\label{bound1}
\|v^\al\|_{C^{r-1}} \le C \|R^\al\|_{C^{r-1}}.
\end{equation}
\end{lemma}

\prova
First of all note that since $f_0\in C^r$ then the bundles $E^s,E^u,TM$ are $C^{r-1}$.
Moreover, since $f_0$ is of class $C^r$, then $F$ is of class $C^{r-1}$, 
and hence $A_\theta(t)$ solving \eqref{aone} depends on a $C^{r-1}$ way on $\theta$ and the derivatives do not
grow with $t$. This implies that we can take derivatives w.r.t. $\theta$ under integral sign in \eqref{int} and thus the bound 
\eqref{bound1} follows.
\EP

To deal with the center direction, since the manifold is normally hyperbolic we can write
$$
v^c(\theta) = DK_0(\theta) w(\theta)\,,
$$
so that the equation for $w$ is
\begin{equation}\label{w}
\om_0\cdot \del_\theta w = (DK_0(\theta))^{-1} R^c(\theta) +\om_n.
\end{equation}
Note that \eqref{w} is the standard cohomology equation appearing in KAM theory. 

In order to solve \eqref{w}, we can
expand $w(\theta)$ in Fourier series
\begin{equation}\label{what}
w(\theta)=\sum_{k\in\ZZZ^d} \hat{w}_k e^{2\pi ik\cdot \theta}\,,
\end{equation}
so that, expanding also
\begin{equation}\label{espo}
g(\theta) := (DK_0(\theta))^{-1} R^c(\theta)   = \sum_{k\in\ZZZ^d} \hat{g}_k e^{2\pi ik\cdot \theta}
\end{equation}
we see that
\eqref{w} reads
\begin{equation}\label{omn}
\begin{aligned}
&\om_n = -\avg{g(\theta)} \\
&\hat{w}_k = \frac{1}{i2\pi \om_0\cdot k}\hat{g}_k,\qquad k\ne0\,.
\end{aligned}
\end{equation}

 In particular in the analytic case, under the hypothesis \eqref{subexp}
we obtain $w(\theta)$ an analytic function defined in a domain $\TTT^d_{\x'}$ for any $\x'<\x$.


Overall we thus obtained the following result.

\begin{theo}\label{analyticnhim}
Suppose that $f_0$ is analytic (resp. $C^\io$), $K_0$ is analytic on $\TTT^d_\x$ (resp. $C^\io$) and $\om_0$ is subexponential Diophantine
(resp. Diophantine).
Then there exists Lindstedt series solving \eqref{embe} to all orders; in particular the coefficients $K_n$ are analytic on $\TTT^d_{\x'}$ for any $\x'<\x$
(resp. $C^\io$). Moreover the Lindstedt series can be made so that \eqref{media} holds; with such normalization the series is unique.
\end{theo}

Note that the $C^r$ case is much trickier to work out. Indeed the solution of the linearized equation looses derivatives
and the composition to justify the derivatives requires justification. Our guess is that one gets $K_n\in C^{r-n\tau}$ and of
course one obtains only finitely many terms in the expansion.

\begin{rmk}\label{liou}
We say that $\om_0$ is Liouville if \eqref{subexp} fails, i.e. there is a sequence $k_n$ with $|k_n|\to\io$ such that $|\om_0\cdot k_n|< c(\e) e^{-\e|k_n|}$.
In this case it is not possible to solve the cohomology equation \eqref{w}, i.e. we cannot find
a Lindstedt series for the solution of the invariance equation. This gives some insight on why the Postnewtonian formalism
(which  is global) fails. Indeed a global theory should work also for Liouville vectors.
\end{rmk}

We finally mention that, under some further technical assumtions (that could possibly be removed), one can apply
the results of \cite{HE1} and get validation for Theorem \ref{analyticnhim}, namely the existence of a true solution nearby.

\zerarcounters 
\section{The reducible case} 
\label{riducibile} 

For linear equations with quasi-periodic coefficients, it is natural to consider linear quasi-periodic changes of variables.
If one is dealing with an equation of the form
\begin{equation}\label{linearqp}
\frac{d}{dt}v = A(\theta+\om_0 t)v
\end{equation}
a change of variables of the form
\begin{equation}\label{cambio}
v(t) = M(\theta+\om_0 t)w(t)
\end{equation}
for some $M$, transforms the equation \eqref{linearqp} into
\begin{equation}\label{cambiata}
\dot{w} = M^{-1}[-\om_0\cdot\del_\theta M + AM](\theta+\om_0 t) w(\theta)\,.
\end{equation}

We say that the equation \eqref{linearqp} is {\it reducible} if it is possible to find
$$
M:\TTT^d_\x\to GL(n,\CCC),\qquad \Lambda\in GL(n,\CCC)
$$
such that
$$
M^{-1}[-\om_0\cdot\del_\theta M + AM](\theta_0+\om t) = \Lambda
$$
so that \eqref{cambiata} has constant coefficients.
Of course, without loss of generality we can look for $\Lambda$ in Jordan normal form.

The question of reducibility has been considered extensively in many papers, both perturbatively \cite{El,DS}
and nonperturbatively \cite{Pu,HY}, in the sense that the smallness condition does not depend on the frequency;
a good survey on the subject can be found in \cite{P}.

What is relevant to us is that, after a change of variables as in \eqref{cambio} we get that \eqref{tengo} becomes
\begin{equation}\label{nova}
\frac{d}{dt}w(t) = \Lambda w + M^{-1}(\theta + \om_0 t) R_\theta+M^{-1}(\theta + \om_0 t)DK_0(\theta+\om_0 t)\om_n
\end{equation}

It is clear that, since
\begin{equation}\label{cappone0}
\frac{d}{dt}K_0(\theta+\om_0 t) = f_0( K_0(\theta+\om_0 t) ,  K_0(\theta+\om_0 t))
\end{equation}
deriving \eqref{cappone0} w.r.t. $\theta$ on both sides we obtain
\begin{equation}\label{automatic}
\frac{d}{dt}DK_0(\theta+\om_0 t) = \big(D_1f_0(K_0(\theta),K_0(\theta))+D_2f_0(K_0(\theta),K_0(\theta))\big)  DK_0(\theta+\om_0 t)\,.
\end{equation}

We can interpret \eqref{automatic} by saying that the vectors $\del_{\theta_i}K_0(\theta)$
are eigenvectors of the linearized equations.

Since $K$ is a torus embedding satisfying \eqref{costante}, there must be $d$ zero-eigenvalues of $\Lambda$. An important assumption
that needs to be made is that there are exactly $d$ zero-eigenvalues of $\Lambda$, while the others
are Diophantine w.r.t. $\om_0$; see Definition \ref{reldio} below.
It is also important to note that the term $R_\theta$ is in the range of $DK_0(\theta)$.
In other words we can split any vector $u$ as
\[
u=\Pi u+ \Pi^\perp u
\]
where $\Pi$ is the projection onto the range of $DK_0$ and $\Pi^\perp$ is the projection onto the complementary space.
Then the invariance equation  restricted to the range of $DK_0$
 takes the form
\begin{equation}\label{hulleq}
\om_0\cdot\del_\theta {\Pi w} = \Pi (M^{-1} {R} + \om_n)\,,
\end{equation}
where, with abuse of notation we are denoting by $w,R$ the corresponding torus embedding.
This is again a standard cohomology equation of the same type of \eqref{w}, so we can solve it by imposing
$$
\om_n =-\avg{(DK_0)^{-1} {R}}
$$
and assuming that $\om_0$ is Diophantine.

On the other hand on the Kernel of $DK_0$ we see that \eqref{hulleq} is equivalent to the system
\begin{equation}\label{system}
\om_0\cdot \del_\theta \Pi^{\perp} {w}_i(\theta) = \mu_i \Pi^\perp {w}_i + \Pi^{\perp}(M^{-1} {R})_i \,,
\end{equation}
where $\mu_{n-d},\ldots\mu_n$ are the non-zero eigenvalues of $\Lambda$.

Note that, because of the previous calculation and the assumption of having no zero-eigenvalues except for the range of $DK_0$,
$\om_n$ does not appear in \eqref{system}.

Similarly to the case of a NHIM, we can now
pass to Fourier series as in \eqref{what} and
we see that \eqref{system} is equivalent to
\begin{equation}\label{fouso}
\begin{aligned}
&2\pi i(\om_0\cdot k ) \hat{w}_k = \mu_i \hat{w}_k + \widehat{M^{-1}R}_{i,k}  \,. \\
\end{aligned}
\end{equation}

This motivates the following definition. 

\begin{defi}\label{reldio}
We say that $\mu_i$ is $(\g,\tau)$-Diophantine w.r.t. $\om_0$ if
\begin{equation}\label{melnikov}
|\mu_i - 2\pi i(\om_0\cdot k) | \ge \frac{\g}{|k|^\tau},\qquad \forall \ k\in\ZZZ^d\setminus\{0\}\,.
\end{equation}
We say that $\mu_i$ is subexponentially Diophantine w.r.t. $\om_0$ if
\begin{equation}\label{submel}
\lim_{|k|\to\io} \frac{1}{|k|}\log|\mu_i - 2\pi i  (\om_0\cdot k)|=0\,.
\end{equation}
\end{defi}

Clearly if $\m_i$ is $(\g,\tau)$-Diophantine w.r.t. $\om_0$ (or subexponentially Diophanitne w.r.t. $\om_0$) we can set 
$$
\hat{w}_k = \frac{1}{\mu_i - 2\pi i  (\om_0\cdot k)} \widehat{M^{-1}R}_{i,k}
$$

It is straightforward to see that if $\mu_i\ne0$ and it is $(\g,\tau)$-Diophantine w.r.t. $\om_0$ we have
$$
\| {w}\|_{\calA_{\x-\de}} \le \g \de^{-(\tau+d)}\| {M^{-1}R}\|_{\calA_\x}
$$

In the case that $\Lambda$ has a non-trivial Jordan block with eigenvalue $\la$ and multiplicity $m$
we obtain a system of equations
$$
\om_0\cdot\del_\theta w^i - \la w^i - w^{i+1}-\ldots - w^m = {R^m},\qquad i=1,\ldots,m
$$
which can be solved recursively starting from order $m$ and going in decreasing order.

Therefore we have the following result.

\begin{theo}\label{teo2}
Assume that
\begin{itemize}

\item $f_0$ is analytic (resp. $\CCCC^\io$).

\item The equation is reducible.

\item The matrix $\Lambda$ has exactly $d$ eigenvalues and the rest of the eigenvalues of $\Lambda$
is subexponentially Diophantine w.r.t. $\om_0$.

\end{itemize}

Then there exists a Lindstedt series solving \eqref{embe} to all orders. The coefficients $K_n$ are  analytic in $\TTT^d_{\x'}$
for all $\x' < \x$ (resp. $\CCCC^\io$)
\end{theo}

\section{Lagrangian tori in the Hamiltonian case}\label{hami}
 
 If we assume that $f_0$ is Hamiltonian\footnote{i.e. $f_0=J\nabla H$ for some function $H:\RRR^{n}\to\RRR$
 with $n$ even,
 and $J$ is the matrix of a 2-form $\Omega(\al,\be)=\laa\al,J\be\raa$, which is symplectic (i.e. $d\Omega=0$
 and non-degenerate)} 
 then in the neighborhood of  an invariant torus there is a very rigid structure that can be used to compute Lindstedt series.
 This structure (called automatic reducibility) was used in \cite{LGJV} to give a computationally efficent proof of the
 KAM theorem; we shall use the automatic reducibility to compute Lindstedt series.
 
 Automatic reducibility has also been found in other systems which preserve geometric structures, such as
 conformally symplectic systems \cite{CCL13} and volume preserving systems \cite{LMJ}.
 The results in this section could also be adapted easily to the other automatically reducible systems.
 
 The key observation is the following result; see \cite{LGJV,Ltutorial}.
 
\begin{lemma}\label{automaticreducibility}
Assume that $f_0:\RRR^n\times\RRR^n\to\RRR^n$ is Hamiltonian and $K_0:\TTT^d\to\RRR^n$ satisfies \eqref{cappone0}. Assume $n=2d$.
Then the $2d\times 2d$ matrix-valued function
$$
M(\theta)=[DK_0(\theta),J^{-1}DK_0(\theta)N(\theta)]
$$
satisfies
\begin{equation}\label{reducibilityhamiltonian}
\om_0\cdot\del_\theta M(\theta) = M(\theta)\begin{pmatrix}
0_d & L(\theta) \cr
0_d & 0_d
\end{pmatrix}
\end{equation}
where $0_d$ denotes the $d\times d$ zero-matrix and $L(\theta)$ is an explicit matrix. By $[\cdot,\cdot]$ we denote the
juxtaposition of two $2d\times d$ matrices to obtain a $2d\times 2d$ matrix.
\end{lemma}

We refer to \cite{LGJV,Ltutorial} for the proof.

Lemma \ref{automaticreducibility} has a very clear geometric meaning. The first columns of $M$ have the same interpretation as
in Section \ref{riducibile}. The last $d$ columns are forced by the preservation of the symplectic
structure.

Using \eqref{reducibilityhamiltonian} we see that, under the change of variables
$$
U=MW
$$
the linearized equation becomes 
$$
\om_0\cdot\del_\theta W = \begin{pmatrix}
0_d & A(\theta) \cr
0_d & 0_d
\end{pmatrix} W + \begin{pmatrix} 0 \cr \om_n\end{pmatrix} + \tilde{R}_n
$$

The reason why $\om_n$ appears only in the second term is that it appears only in $DK_0(\theta)\om_n$.

Again we obtain $\om_n$ so that the second term has a solution using the theory of constant differential equations.
Then the second component is determined up to an additive constant (the constant is uniquely determined if we impose
for instance \eqref{media}).

To summarize, we obtained the following result.

\begin{theo}\label{teo3-2}
Assume that $n=2d$ and
 $f_0$ is an analytic (resp. $\CCCC^\io$) Hamiltonian vector field.
Then there exists a Lindstedt series solving \eqref{embe} to all orders. The coefficients $K_n$ are  analytic in $\TTT^d_{\x'}$
for all $\x' < \x$ (resp. $\CCCC^\io$).
\end{theo}

\section{Limit cycles and isochrones} 
\label{sec:limitcycle}

We now consider the case in which, for $\e=0$, \eqref{gene}
or \eqref{vabbeh}) is a 
two dimensional ODE, which admits a limit cycle. 
This model is very common in applications in electronics, where 
the classical models of oscillators are limit cycles, but for 
fast electronics it is useful to include the delay. 

Besides the limit cycle, it is useful to consider the solutions that 
converge exponentially to it. They were called isochones in 
\cite{Winfree} which explained their 
physical and biological relevance. Their
relation to stable manifolds was pointed out in 
\cite{Guckenheimer}. 

In this section, we will show that there are Lindstedt series 
both for the limit cycles and for the isochrones.  Furthemore 
we will mention that this fits very well with the recent developments 
in a-posteriori theorems \cite{YangGL19, GimenoYL19} and that, 
in this case, we can prove that the series are asymptotic in 
a very strong sense. See Theorem~\ref{teo4}. 

We also note that in this case we will develop a method to compute
the Linstedt series in a much faster way. Each step of the algorithm
will double the number of computed terms. This is in contrast with the 
methods discussed before, in which one step of 
the algorithm produced only one more term in the expansion. We may informally
describe this method as overloading the Newton method to power series.

A convenient  starting point for our 
analysis is the result  in  \cite{HL}
that in a neighbohood of the limit cycle, 
there is an embedding
\begin{equation}\label{vuddoppione}
W:\TTT^1 \times\RRR\to \RRR^2
\end{equation}
so that  for every $\theta_0 \in  \TTT$ and $ s_0 \in \RRR$ with $|s_0|\ll1$,
\begin{equation} \label{solutionform}
y(t)=W(\theta_0+\om_0 t, s_0e^{\la t})
\end{equation} 
 is a solution of 
the equation of $\e = 0$.  What we will do is to seek 
to modify the $W$, $\om$, $\la$ so that \eqref{solutionform} is a solution of the delay equation.

Note that in this case, we are not looking 
for a torus embedding as in \eqref{embe} or 
\eqref{embe2}, but we are also including the exponentially 
converging orbits.  Hence, there are 
two parameters to be found,  $\om$, the frequency of the torus
and $\la$, the exponential factor of convergence.

Finding solutions of the delay equation  \eqref{gene} or \eqref{vabbeh}
of the form  \eqref{solutionform} is  equivalent to 
finding $W, \om, \la$ satisfying
\begin{equation}\label{embeciclo}
(\om\cdot\del_\theta + s\la \del_s) W(\theta,s) = f_\e (W(\theta,s),W(\theta - \e \om r(W(\theta,s)), se^{-\e \la r(W(\theta,s))})).
\end{equation}
or
\begin{equation}\label{embeciclo2}
(\om\cdot\del_\theta + s\la \del_s) W(\theta,s) = f_\e (W(\theta,s),\e W(\theta -  \om r(W(\theta,s)), se^{-\la r(W(\theta,s))})).
\end{equation}
respectively.

Again we look for a solution $( W(\theta,s), \om, \la) $ 
of \eqref{embeciclo} as a formal power series, i.e.
\begin{equation}\label{espando}
\begin{aligned}
\la = \sum_{j\ge0}&\e^j \la_j,\qquad \om = \sum_{j\ge0}\e^j \om_j  \\
&W(\theta,s) = \sum_{j\ge 0 } \e^j W_j(\theta,s)\,,
\end{aligned}
\end{equation}

The case is a particular case of the reducibility case, 
so we could get the series using the methods in 
Section~\ref{riducibile}.   
In this section, however, we want to describe a different algorithm 
that is based on a Newton method and is quadratically convergent. 
When applied to the problem of Lindstedt series, we see 
that the method will double the number of coefficients that we have
computed at every step (the step will be more complicated than
in the order by order method). In this paper, 
we will not perform a comparision of 
the computational cost of the Newton method and the order by order method. 
In \cite{HL} such comparisons are performed in the ODE case.

Of course the Meta-Lemma \ref{formal} applies in a slightly different form also in this case, so we need to show
that we can solve the linearized equation. Indeed, set
\begin{equation}\label{ellone}
L_{\om,\la} = \begin{pmatrix} \om \cr \la s \end{pmatrix},
\end{equation}
 denote
\begin{equation}\label{D}
DWL_{\om ,\la} = (\om\cdot\del_\theta + s\la \del_s) W(\theta,s) \\
\end{equation}
and
\begin{equation}\label{F}
\begin{aligned}
&F\circ W = f_\e (W(\theta,s),W(\theta - \e \om r(W(\theta,s)), se^{-\e \la r(W(\theta,s))})),\qquad\mbox{ or}\\
&F\circ W = f_\e (W(\theta,s),\e W(\theta -  \om r(W(\theta,s)), se^{-\la r(W(\theta,s))}))
\end{aligned}
\end{equation}
for \eqref{embeciclo} and \eqref{embeciclo2} respectively,
and assume that we have an approximate solution $(W,\om,\la)$ of \eqref{embeciclo}, i.e. such that 
\begin{equation}\label{newt}
\begin{aligned}
DWL_{\om,\la} = F \circ W  + E
\end{aligned}
\end{equation}
for some small $E$.
Thus for the Newton scheme we need to find an better approximation $(W+\Delta,\om+\al,\la+\be)$ such that the correction $(\Delta,\al,\be)$
elimitates the error $E$ at the linear approximation. This means that indeed we need to solve the linearized equation
\begin{equation}\label{lin}
D\Delta L_{\om,\la} + DWL_{\al,\be } = (DF\circ W) \Delta+E\,.
\end{equation}

Note that differentiating \eqref{newt} we get
\begin{equation}\label{quasi}
D^2WL_{\om,\la} +DWDL_{\om,\la}= (DF\circ W) DW + DE
\end{equation}
so, since the operator $DW$ is invertible, the idea is to look for $\Delta$ of the form
\begin{equation}\label{figata}
\Delta = DWA\,.
\end{equation}
Substituting \eqref{figata} into \eqref{lin} and rearranging we get
\begin{equation}\label{quasi2}
D^2WA L_{\om,\la} + DWDAL_{\om,\la} - (DF\circ W) DWA= - DWL_{\al,\be }  +E\,.
\end{equation}
Now, since $D^2W AL_{\om,\la} = D^2WL_{\om,\la}A$, and using \eqref{quasi2} we obtain
$$
(DE- DWDL_{\om,\la}) A + DWDAL_{\om,\la} = - DWL_{\al,\be }  +E\,.
$$
Due to the fact that the term $DEA$ is ``quadratically small'', we may drop it so that the equation for $A$ becomes
\begin{equation}\label{quasinewt}
- DWDL_{\om,\la} A + DWDAL_{\om,\la} = - DWL_{\al,\be }  +E.
\end{equation}
This last step is called \emph{quasi-Newton step} in the literature; see for instance \cite{HL} and references therein. If 
we now multiply by $DW^{-1}$ we see that \eqref{quasinewt} reduces to
$$
-DL_{\om,\la} A + DAL_{\om,\la} = - L_{\al,\be }  +\tilde{E},\qquad \tilde{E}:=DW^{-1}E,
$$
which in components $A=(A_1,A_2)$, $\tilde{E}=(\tilde{E}_1,\tilde{E}_2)$, takes the form
\begin{equation}\label{eccola}
\begin{aligned}
&(\om\cdot \del_\theta+s\la \del_s)A_1 + \al = \tilde{E}_1 \\
&(\om\cdot \del_\theta+ s\la\del_s-\la)A_2 + \be s = \tilde{E}_2\,,
\end{aligned}
\end{equation}
i.e. it is a linear equation with constant coefficients.
Equations like \eqref{eccola} were studied in \cite{HL} with two methods. Here we follow the analysis based on
power series. Indeed, expanding
\begin{equation}\label{fouciclo}
\begin{aligned}
&A_h(\theta,s)=\sum_{k\in\ZZZ^d} \sum_{p\ge0}\hat{A}_{h,k}^{(p)}s^p e^{2\pi ik\theta},\qquad h=1,2 \\
&\tilde{E}_h(\theta,s)= \sum_{k\in\ZZZ^d}\sum_{p\ge0} \hat{\tilde{E}}_{h,k}^{(p)}s^p e^{2\pi ik\theta},\qquad h=1,2 \\
\end{aligned}
\end{equation}
we see that \eqref{eccola} takes the form 
\begin{equation}\label{eccolaf}
\begin{aligned}
&(i2\pi \om\cdot k+\la p)\hat A_{1,k}^{(p)}  = \hat{\tilde{E}}_{1,k}^{(p)}\qquad p\ge1 \\
&(i2\pi \om\cdot k+ \la p-\la)\hat A_{2,k}^{(p)}  =\hat{ \tilde{E}}_{2,k}^{(p)}\qquad p=0,\ \  p\ge2\,,
\end{aligned}
\end{equation}
and
\begin{equation}\label{zeroo}
\begin{aligned}
&(i2\pi \om\cdot k)\hat A_{1,k}^{(0)} + \al  = \hat{\tilde{E}}_{1,k}^{(0)} \\
&(i2\pi \om\cdot k)\hat A_{2,k}^{(1)}  + \be  = \hat{\tilde{E}}_{2,k}^{(1)}
\end{aligned}
\end{equation}

Thus we can fix
\begin{equation}\label{para}
\begin{aligned}
&\al=\hat{\tilde{E}}_{1,0}^{(0)},\qquad \be =\hat{\tilde{E}}_{2,0}^{(1)},\\
&\hat A_{1,k}^{(0)}  = \frac{\hat{\tilde{E}}_{1,k}^{(0)}}{(i2\pi \om\cdot k)} \qquad 
\hat A_{2,k}^{(1)}   =\frac{ \hat{\tilde{E}}_{2,k}^{(1)}}{(i2\pi \om\cdot k)},\qquad k\ne0
\end{aligned}
\end{equation}
and
\begin{equation}\label{glia}
\hat{A}_{1,k}^{(p)} =\frac{\hat{\tilde{E}}_{1,k}^{(p)}}{i2\pi \om\cdot k+\la p} \qquad \hat A_{2,k}^{(p)} = \frac{\hat{\tilde{E}}_{2,0}^{(p)}}{i2\pi \om\cdot k+\la (p-1)}\,.
\end{equation}
Since $\la$ and $p$ are both real, then a subexponential Diophantine $\om_0$ ensures $A_{1}^{(p)}$, $A_2^{(p)}$ to be analytic functions of $\theta$. Precisely we proved
the following result.

\begin{theo}\label{teo3}
Assume that $f_0$ is analytic (resp. $\CCCC^\io$).
Then there exists Lindstedt series
\[
W =\sum_n \e^n\sum_{p\ge0} s^pW_j^{(p)}, \quad \om = \sum_n \e^n \om_n  \quad \la = \sum_n \e^n \la_n
\]
  solving \eqref{embeciclo} (resp. \eqref{embeciclo2}) to all orders. The coefficients $W_j^{(p)}$ are  analytic in $\TTT^d_{\x'}$
for $\x' <\x$ (resp. $\CCCC^\io$)
\end{theo}

We also mention that in this case there is an a-posteriori theory 
developed in \cite{YangGL19}, which takes as principal input 
the fact that there are approximate solutions that solve very approximately 
the equation \eqref{embeciclo} and conclude that there 
are true solutions. 

Since the main conclusions of Theorem~\ref{teo3} are precisely that 
we can construct series that satisfy \eqref{embeciclo} very accuratey, 
we can put together 
Theorem~\ref{teo3} and the results of 
\cite{YangGL19} and  we obtain the following.
   
\begin{theo}\label{teo4}
In the assumptipions of Theorem~\ref{teo3}, 
we can find solutions $W_\e, \om_\e \la_\e $ of the equation
\eqref{embeciclo}.  These $W_\e$ are finitely differentiable functions 
for any $\e > 0$. 

Furthermore, there exists a function $r(\e)$,
with $\lim_{\e \to 0} r(\e) = \infty $ in such a
way that for all $N$,  there exist numbers $C_N$ such that 
\[
\begin{split} 
&  \| W_\e - W^{[\le N]}_\e \|_{C^{r(\e)}} \le C_N \e^{N+1} \\
&  | \om_\e - \om^{[\le N]}_\e | \le C_N \e^{N+1} \\
&  | \la_\e - \la^{[\le N]}_\e | \le C_N \e^{N+1} \\
\end{split} 
\]
\end{theo}

Note that the conclusions are slightly stronger than the 
usual definition of asymptotic expansions since we conclude 
that the approximation is happening in stronger norms as 
 $\e$ goes to zero.

\section{Systems with more delays and the Electrodynamics case}
\label{sec:electrodynamics}

We proved that it is possible to find solutions (in the sense of formal power series) to SDDE equations of the form \eqref{gene} or \eqref{vabbeh}
 in various setting
when $\ell=1$. It is however clear that with a slight modification of the discussions above we could cover the case $\ell\ge2$.
Indeed the only difference is that the vector field $f_\e$ depends on $\ell+1$ arguments instead of only two,
so it suffices to replace the operator 
$$
 D_1 f_0(K_0(\theta),\ldots, K_0(\theta)) +  D_2 f_0(K_0(\theta),\ldots, K_0(\theta))
$$
 with
$$
\sum_{p=1}^{\ell+1} D_p f_0(K_0(\theta),\ldots, K_0(\theta))
$$
where $D_p$ denotes the derivative w.r.t. the $p$-th argument.

We now discuss the physical case \eqref{dynamics}. 
We start by rewriting \eqref{dynamics} as a dynamical system, i.e.
\begin{equation}\label{dinamico}
\begin{aligned}
&\dot{x}_i = v_i \\
&\dot{v}_i = {G}{M_i(v_i)^{-1}}\sum_{j\ne i} \frac{q_i q_j  (x_i(t) - x_j(t-\e|x_i(t)-x_j(t)|+O(\e^2) ))}{|x_i(t) - x_j(t-\e|x_i(t)-x_j(t)|+O(\e^2))|^3}
\end{aligned}
\end{equation}
where we denoted $\e=1/c$ and we also exploited the expansion \eqref{tauexpansion}. We then look for a torus embedding
\begin{equation}\label{torone}
K:\TTT^d\to\RRR^{6N}
\end{equation}
satisfying an equation of the form
\begin{equation}\label{evero}
\om\cdot \del_\theta K(\theta) = F(K(\theta),K(\theta - \e\om r(K(\theta)) + O(\e^2))).
\end{equation}
By Remark \ref{ezero} we see that if $d=3N$ we are essentially in the same situation as in Case 3, so we can apply the results of Section \ref{hami}.

\appendix

\zerarcounters
\section{Solutions of cohomology equations with frequency given by formal power series}

In all the cases studied in the present paper, the frequencies are given by power series in $\e$. However
we required \eqref{dio} or \eqref{subexp} only for the first summand of the series defining $\om$.
Indeed the following is true.

\begin{lemma}\label{a1}
Let
\begin{equation}\label{ome}
\om=\om_\e = \sum_{j\ge 0}\e^j \om_j
\end{equation}
be an $\RRR^d$-valued formal power series. Let
\begin{equation}\label{eta}
\h=\sum_{j\ge0}\e^j \h_j
\end{equation}
be an $\calA_\x$-valued formal series (recall \eqref{ax}), i.e. $\h_j\in\calA_\x$ for all $j\ge0$.
Assume that $\om_0$ is subexponential Diophantine (recall \eqref{subexp}). Then for every $\de>0$
there is a unique $\calA_{\x-\de}$-valued formal power series
\begin{equation}\label{fi}
\f=\sum_{j\ge0}\e^j\f_j
\end{equation}
solving
\begin{equation}\label{co}
\om\cdot\del_\theta \f = \h
\end{equation}
in the sense of power series. Moreover the solution $\f$ is unique if we impose
\begin{equation}\label{av}
\frac{1}{(2\pi)^d}\int_{\TTT^d} \f_j(\theta)d\theta = 0 \,,\qquad j\ge0\,.
\end{equation}
\end{lemma}

The key observation to prove Lemma \ref{a1} is the following (very well known) result.

\begin{prop}\label{barbatrucco}
If $\om$ is of the form \eqref{ome} and $\om_0$ satisfies \eqref{subexp}, then given any $\al\in\calA_\x$, for any $\x'<\x$ there is a
solution to
\begin{equation}\label{be}
\om\cdot\del_\theta \be = \al\,.
\end{equation}
\end{prop}

\prova
If we Fourier-expand
\[
\al(\theta)=\sum_{k\in\ZZZ^d}\hat{\al}_k e^{2\pi i k\cdot \theta},\qquad
\be(\theta)=\sum_{k\in\ZZZ^d}\hat{\be}_k e^{2\pi i k\cdot \theta}\,,
\]
then we see that \eqref{be} is equivalent to 
\[
(2\pi i \om\cdot k)\hat{\be}_k = \hat{\al}_k.
\]
If $\om_0$ satisfies \eqref{subexp} then
\[
|\om\cdot k|^{-1} \le C e^{(\x-\x')|k|/2}\,.
\]

By Cauchy estimates we have
\[
|\hat{\al}_k| \le e^{-2\pi\x|k|}\|\al\|_\x\,,
\]
and hence
\[
\|\be\|_{\x'} \le \sum_{k\in\ZZZ^d} e^{-2\pi\x|k|}\|\al\|_\x Ce^{(\x-\x')|k|/2} \le \tilde{C}\|\al\|_\x \sum_{k\in\ZZZ^d} e^{-|k|(\x-\x')/2}\,,
\]
for some constant $\tilde{C}$,
so the assertion follows.
\EP

We are now ready to prove Lemma \ref{a1}

\prova {\it (Lemma \ref{a1})}
We can rewrite \eqref{co} as
\begin{equation}\label{formalco}
\om_0\cdot\del_\theta \f_n = \h_n - \sum_{j=1}^n \om_j\cdot\del_\theta \f_{n-j}\,,
\end{equation}
hence we can use recursively Proposition \ref{barbatrucco} to find
\[
\f_n \in\calA_{\x -(1-2^{-n})\de}
\]
so the assertion follows.
\EP


\def\cprime{$'$}

\end{document}